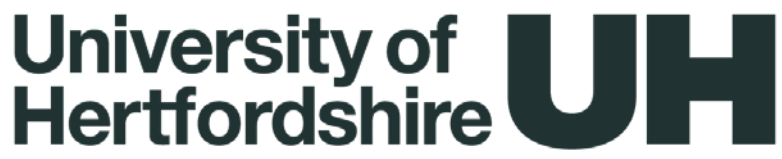

School of Physics, Engineering
and Computer Science

# Multimodal Brain Tumour Classification Using Feature Fusion

**Wajih ul Islam[1*], Muhammad Yaqoob[1], Javed Ali Khan[1] and Volker Steuber[1]**
[1]*School of Physics, Engineering and Computer Science, University of Hertfordshire, UK*
*corresponding author: w.u.islam@herts.ac.uk

**Clinicians diagnose brain tumors by synthesizing patient symptoms, medical history, and quantitative imaging data from modalities such as MRI and CT scans into a unified clinical judgement. However, most deep learning models rely on MRI/CT images alone, failing to replicate the clinicians' multimodal reasoning. We explore a two-branch multimodal network combining raw MRI scans with 91 extracted radiomic features (intensity, texture, shape, and boundary descriptors) to classify brain tumors into glioma, meningioma, pituitary, and no-tumor. A pre-trained CNN backbone encodes the image stream whereas a dedicated MLP encodes the radiomic stream. Both streams are fused via concatenation, gated, or bidirectional cross-modal attention strategies. Across nine experimental runs on a balanced 7,200-image dataset, all multimodal configurations outperform unimodal baselines with gated fusion achieving the best accuracy of 96.13%.**

## Introduction

In clinical practice, medical practitioners diagnose brain tumor by integrating multiple information sources i.e., symptoms, history, laboratory results, and imaging such as MRI. However, the majority of deep learning models for brain tumor classification operate on image data alone[1]. This unimodal approach to deep learning fails to replicate the multimodal reasoning in the real diagnostic workflows. To address this, we explore a two-branch multimodal framework that combines raw MRI scans with 91 radiomic features extracted per image to classify brain tumors into four categories: glioma, meningioma, pituitary, and no-tumor. A pre-trained CNN backbone (EfficientNet-B3, ResNet-50, or DenseNet-121) encodes the image stream whereas a dedicated MLP encodes the radiomic stream, with both projected to a shared 256-dimensional embedding space and fused via one of three strategies: concatenation, gated fusion with learned per-dimension modality weights, and bidirectional cross-modal attention. We performed a grid of nine experimental runs to investigate how backbone capacity and fusion strategy impact the classification performance.

## Dataset

The publicly available dataset comprises 7,200 brain MRI images distributed across four tumor categories: glioma, meningioma, pituitary tumor, and no-tumor. The dataset is perfectly balanced, with 1,600 images per class in the training set and 400 per class in the test set. The dataset was compiled from multiple publicly available MRI repositories, including the Sartaj dataset, BRaTS MRI images, and other open access sources [2,3], introducing diversity in scanner protocols, acquisition settings, and patient populations. The predefined training and testing split provided with the dataset was used in this study. Since patient identifiers were not available split should be considered imagel evel rather than patient level, and possible patient overlap between training and testing subsets cannot be completely excluded.This limitation is acknowledged in the present preliminary study.

## Experimental Setup

In the preliminary setup, radiomic features are extracted from each MRI scan and processed via a dedicated MLP to produce a radiomic embedding, and the raw MRI scan is passed through a pre-trained CNN backbone to produce a visual embedding; both are subsequently fused for classification. To extract the radiomic features, the MRI scan is resized to 224×224 pixels and passed through a three-stage unsupervised segmentation pipeline to extract three spatial regions: a brain mask extracted via CLAHE contrast enhancement and a double Otsu threshold, a tumor ROI identified by combining a global top-20% intensity threshold with a local adaptive threshold followed by morphological refinement, and a 15-pixel peritumoral ring surrounding the tumor core.

From these regions, 91 radiomic features are extracted across ten families: first-order intensity statistics (15), shape descriptors (12), GLCM texture (12), GLRLM run-length (8), NGTDM neighborhood tone (5), GLDM grey-level dependence (8), gradient and HOG features (12), peritumoral first-order (8), peritumoral GLCM (6), and targeted boundary/heterogeneity/fractal features (5) [4]. These numerical features are exported per image with a mask reliability flag into a CSV file.

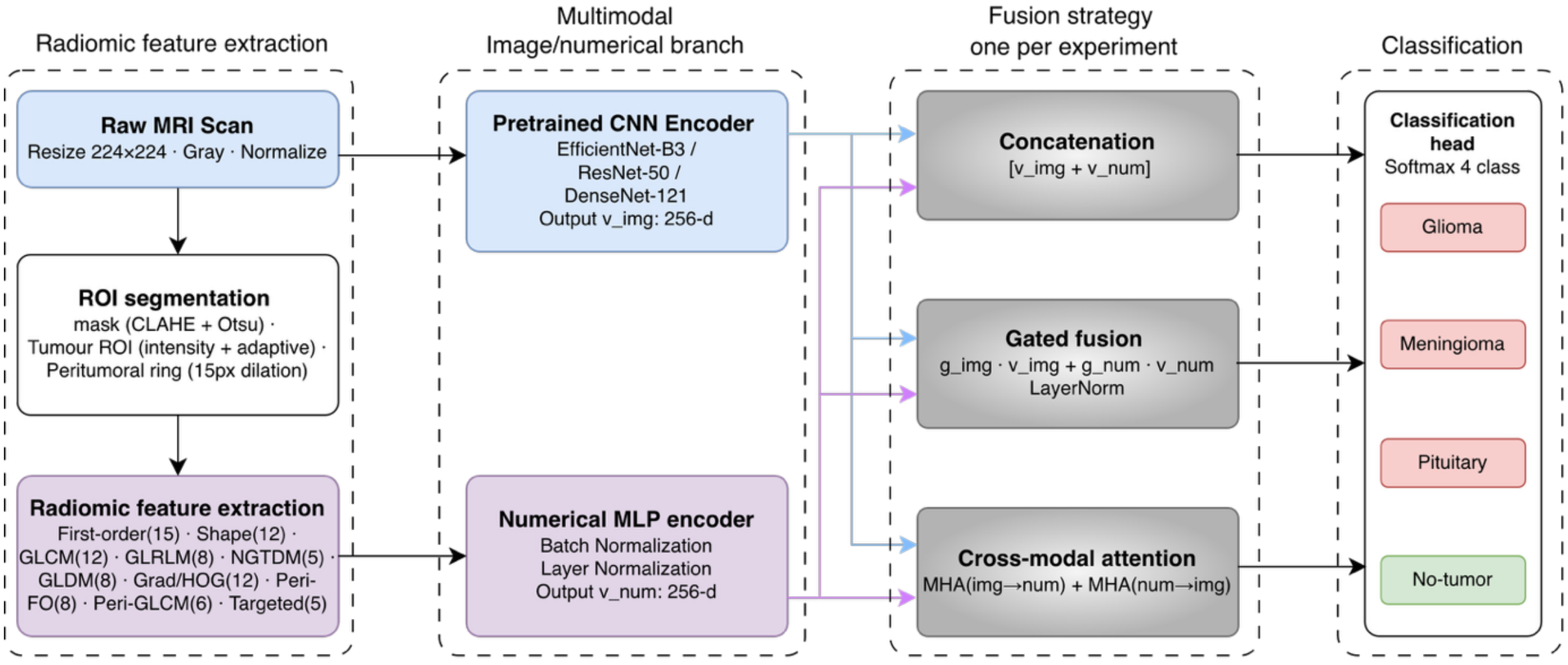

*Fig. 1. Preliminary experiment on multimodal brain tumor classification. First ROI is segmented and 91 radiomic features are extracted. Then multimodal image/numerical branches are fused via concatenation, gated, or cross-modal attention for four-class tumor prediction.*

The obtained radiomic feature vectors are processed via a numerical MLP encoder whereas the original MRI images pass through the CNN encoder. One of three pre-trained CNN backbones: EfficientNet-B3, ResNet-50, or DenseNet-121 are evaluated based on their complementary architectural designs and proven performance in medical image classification tasks [5]. The CNN outputs a 256-dimensional embedding via a dropout, linear, Layer Norm head. The radiomic features are independently encoded by a three-layer MLP (Batch Norm, Linear(128) , Linear(256), Layer Norm) to a matching 256-d embedding. The two embeddings are then fused using one of three strategies evaluated across separate runs: simple concatenation, gated fusion with dual learned sigmoid gates, or bidirectional cross-modal attention with residual connections (Fig. 1). During training the backbone is frozen for 8 epochs then fine-tuning the last three blocks for 12 epochs with weighted sampling, label smoothing, gradient clipping, and mixed precision. We performed an experiment grid of 9 runs (3 backbones × 3 fusion strategies) to evaluate the utility of multimodal fusion.

## Results and discussion

The individual CNNs established a baseline, with DenseNet-121 achieving the highest single-modality accuracy of 95.50%, slightly outperforming ResNet-50 (95.37%) and EfficientNet-B3 (94.94%), with meningioma consistently being the most challenging class across all three backbones due to its lower recall relative to the other classes (Table 1). When the radiomic branch was introduced, all multimodal configurations consistently outperformed their corresponding unimodal baselines across nine experimental runs demonstrating the robustness of multimodal feature fusion across different backbone architectures and fusion strategies. The best performing configurations were ResNet-50 + Gated fusion and DenseNet-121 + Gated fusion both achieving 96.13% followed closely by DenseNet-121 + Attention (96.06%) and DenseNet-121 + Concatenation (95.81%). Across fusion strategies gated fusion demonstrated the most consistent gains (Table 1). Moreover no-tumour and pituitary classes benefited most from multimodal fusion with near perfect recall across nearly all configurations. Meningioma remained the hardest class to classify in both unimodal and multimodal settings due to its heterogeneous appearance and morphological overlap with other tumour types. The proposed framework achieves competitive classification performance while integrating handcrafted radiomic descriptors with deep CNN image embeddings through multimodal fusion. Comparison with recent literature further highlights the effectiveness of the proposed approach. Aziz et al. (2024) reported an accuracy of 96.9% using a DenseNet-based MRI classification framework, whereas the proposed multimodal CNN + radiomics fusion framework achieved 96.13% accuracy while incorporating both deep image embeddings and handcrafted radiomic descriptors.

*Table 1. Per-class F1-scores and overall accuracy for unimodal (MRI only) and multimodal (image + 91 radiomic features) across three CNN backbones and three fusion strategies.*

| Model | Type | Glioma | Meningioma | No-tumour | Pituitary | Accuracy |
|---|---|---|---|---|---|---|
| **EfficientNet-B3** | Uni | 0.897 | 0.933 | 0.976 | 0.988 | 94.94% |
| + Concatenation | Multi | 0.913 | 0.932 | 0.993 | 0.988 | 95.69% |
| + Gated | Multi | 0.903 | 0.932 | 0.99 | 0.982 | 95.25% |
| + Attention | Multi | 0.91 | 0.932 | 0.999 | 0.986 | 95.75% |
| **ResNet-50** | Uni | 0.91 | 0.943 | 0.962 | 0.996 | 95.37% |
| + Concatenation | Multi | 0.913 | 0.941 | 0.979 | 0.991 | 95.69% |
| + Gated | Multi | 0.924 | 0.942 | 0.986 | 0.991 | 96.13% |
| + Attention | Multi | 0.915 | 0.935 | 0.994 | 0.984 | 95.75% |
| **DenseNet-121** | Uni | 0.906 | 0.944 | 0.971 | 0.995 | 95.5% |
| + Concatenation | Multi | 0.915 | 0.931 | 1 | 0.985 | 95.81% |
| + Gated | Multi | 0.922 | 0.946 | 0.98 | 0.995 | 96.13% |
| + Attention | Multi | 0.918 | 0.932 | 0.998 | 0.994 | 96.06% |

*Table 2. Comparison of the proposed multimodal framework with recent brain tumor MRI classification studies*.

| Related work | Model | Accuracy |
|---|---|---|
| Rasheed et al. (2023) | CNN with image enhancement | 95.56% |
| Aziz et al. (2024) | Dense Net MRI classification | 96.9% |
| Proposed work | CNN + radiomics fusion | 96.13% |

## Conclusion

This preliminary work proposes a multimodal framework for brain tumor classification that combines CNN-based image encoding with handcrafted radiomic features across three backbone architectures and three fusion strategies. We demonstrate that multimodal fusion consistently outperformed unimodal image only baselines across all nine experimental configurations. The best performing models were ResNet-50 and DenseNet-121 with gated fusion, both achieving an accuracy of 96.13%. The current multimodal framework uses two complementary representations derived from the same MRI source: deep image embeddings and handcrafted radiomic descriptors. However, it is important to acknowledge that the observed improvement over the unimodal baselines is modest, which is largely attributable to the fact that both branches derive their information from the same MRI image. Since the radiomic features are extracted from the same source signal as the CNN input, the degree of true complementarity between modalities is limited. Future work will include a radiomics only baseline and feature family ablation studies to quantify the contribution of individual radiomic descriptors. In the future, we plan to evaluate the multimodal deep models by replacing the radiomic branch with independent clinical data sources, e.g. structured clinical reports, laboratory results, or patient history, which would introduce information entirely absent from the image domain and more accurately replicate the multimodal reasoning of clinical diagnosis.